\documentclass[conference,letterpaper]{IEEEtran}

\usepackage[utf8]{inputenc} 
\usepackage[T1]{fontenc}
\usepackage{url}
\usepackage{ifthen}
\usepackage{cite}
\usepackage[cmex10]{amsmath} 
\usepackage{cite}
\usepackage{amsmath,amssymb,amsfonts}
\usepackage{amsthm}
\usepackage{algorithmic}
\usepackage{graphicx}
\usepackage{textcomp}
\usepackage{xcolor}
\usepackage{bbm}
\usepackage{mathrsfs}
\usepackage{algorithm}
\usepackage{algorithmic}
\usepackage{booktabs}
\usepackage{balance} 
\usepackage{subcaption}
\usepackage{graphicx}
\newtheorem{theorem}{Theorem}[section]
\newtheorem{proposition}{Proposition}[section]

\begin{document}
\title{ Secure Inference for Vertically Partitioned Data Using Multiparty Homomorphic Encryption} 



\author{%
}
\author{%
  \IEEEauthorblockN{Shuangyi Chen\IEEEauthorrefmark{1},
                    Yue Ju\IEEEauthorrefmark{2},
                    Zhongwen Zhu\IEEEauthorrefmark{2},
                    and Ashish Khisti \IEEEauthorrefmark{1}}
  \IEEEauthorblockA{\IEEEauthorrefmark{1}%
                    University of Toronto, 
                    \{shuangyi.chen@mail.utoronto.ca, akhisti@ece.utoronto.ca\}}
  \IEEEauthorblockA{\IEEEauthorrefmark{2}%
                    Ericsson-GAIA Montréal,
                    \{yue.ju, zhongwen.zhu\}@ericsson.com}
}

\maketitle

\begin{abstract}
    We propose a secure inference protocol for a distributed setting involving a single server node and multiple client nodes. We assume that  the observed data vector is partitioned across multiple client nodes while the deep learning model is located at the server node. Each client node is required to encrypt its portion of the data vector  and transmit the resulting ciphertext to the server node. The server node is required to collect the ciphertexts and perform inference in the encrypted domain. We demonstrate an application of multi-party homomorphic encryption (MPHE) to satisfy these requirements. We propose a  packing scheme, that enables  the server to form the ciphertext of the complete data by aggregating the ciphertext of data subsets encrypted using MPHE.
    While our proposed protocol builds upon prior horizontal federated training protocol~\cite{sav2020poseidon}, we focus on the inference for vertically partitioned data and avoid the transmission of (encrypted) model weights from the server node to the client nodes.
\end{abstract}

\section{Introduction}
Prediction-as-a-service refers to a scenario that a service provider deploys its neural network models on the cloud and allows users to upload their input to obtain the inference result. In the setting of prediction-as-a-service, a secure inference protocol enables the service provider and users securely interact to evaluate the model while preserving the privacy of both the model and the input. Cryptographic tools such as Homomorphic Encryption (HE) which allows the computation on the encrypted data, have been widely applied in this area. A pure HE-based protocol usually protects the model parameters but also the architecture of the model. CryptoNets \cite{gilad2016cryptonets} was the earliest work to use HE on neural network inference. Inspired by CryptoNets, many works such as CryptoDL \cite{cryptodl} and CHET \cite{dathathri2019chet} focus on the optimization of the approximated polynomials for non-linear functions which are not supported by HE computation. Such pure HE-based secure inference protocols have been well developed when the input to the model is provided by a single party. However, those existing protocols do not immediately generalize to a setting where the inputs to the model are vertically distributed among multiple parties. On the other hand, existing HE-based vertical federated learning (VFL) protocols \cite{hardy2017private, yang2019quasi, xu2021fedv, 10206955} that naturally support the inference have several limitations. Firstly, some \cite{hardy2017private, yang2019quasi} are 2-party protocols and cannot be extended to multiparty setting. Furthermore, some \cite{xu2021fedv} cannot protect the confidentiality of both the input and the model. Last but not the least, some \cite{xu2021fedv, 10206955} can be applied only to limited models, typically simple ML models. Those limitations make the VFL protocols not the best choice for building a secure inference protocol with vertical partitioned data as input. 

\begin{figure}[h]
    \centering
    \includegraphics[scale=0.36]{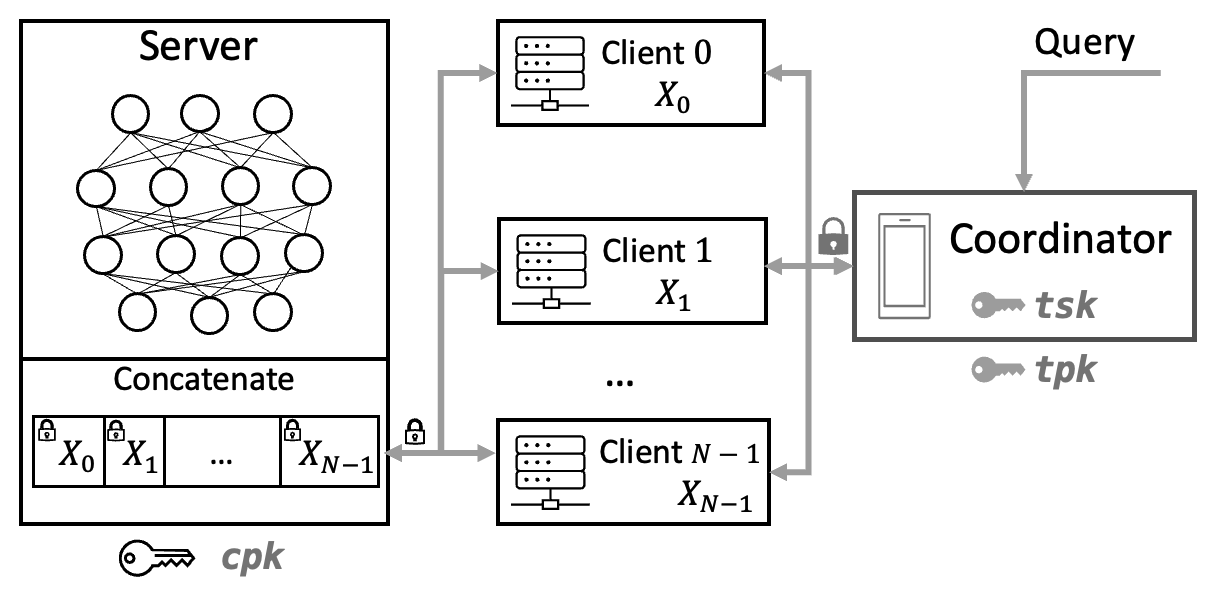} 
    \caption{System setup}
    \label{fig:setting}
\end{figure}

Considering the mentioned limitations, we propose an end-to-end N-party secure inference protocol for vertical partitioned data for simple linear models but also for neural networks, as shown in Figure~\ref{fig:setting}. Our protocol ensures the confidentiality of local data and models, maintaining robustness under a passive adversarial model and against collusions involving up to $N-1$ parties. 
In our protocol, the client nodes encrypt their respective portion of data vector using MPHE and transmit to the server, where the inference is executed in the encrypted domain. Our main contributions are as follows:
\begin{enumerate}
    \item \textbf{Novel packing scheme}: Our proposed approach named \textsf{V-Pack} enables the client to pre-process its subset features with only the knowledge of its own feature index and the structure information of the model and enables the server to form a ciphertext of a complete sample from clients' ciphertext of its own packed features without knowing clients' identities.
    \item {\bf  Implementation and Analysis}: Our proposed protocol is also secure against a dishonest majority of client nodes. We also implement our protocol and present experimental results illustrating the computational and communication costs associated with our  implementation.
\end{enumerate}

Our protocol employs cryptographic neural network operations introduced by a prior work named POSEIDON \cite{sav2020poseidon}, an N-party horizontal federated training system relying on Multiparty Homomorphic Encryption (MPHE) \cite{MPHE}. 

\section{Preliminaries}
\subsection{Multiparty Homomorphic Encryption} \label{sec:MPHE}
We rely on Multiparty Homomorphic Encryption scheme (MPHE) to build the secure inference protocol for vertical partitioned data. Multiparty Homomorphic Encryption scheme \cite{MPHE} enables multiple distribute parties, each with an input in private, to collaboratively compute a function of those inputs without leaking any party's input. In this scheme, a common public key $cpk$ collectively generated by parties is known to all parties, while the corresponding secret key $csk$ is distributed among parties. The decryption needs the participation of all the parties that contribute to the common public key. The security of this scheme is based on the assumed hardness of the ring learning with errors (RLWE) problem \cite{lyubashevsky2010ideal}. We rely on Cheon-Kim-Kim-Song (CKKS) scheme \cite{cheon2017homomorphic}, which is an approximate HE scheme. CKKS supports floating point arithmetic, which makes it the most suitable option for constructing the encrypted neural network operations. We define the ciphertext space as $R_{Q_L}= \mathbbm{Z}_{Q_L}[X]/(X^{\mathscr{N}}+1)$ where $\mathscr{N}$ is the dimension of the cyclotomic polynomial ring, a power-of-two integer. $Q_L = \prod_{i=0}^{L}q_i$ is the ciphertext modulus at the initial level $L$ where each $q_i$ is a unique prime. Below we introduce the main functions used in the protocol. Consider $N$ parties want to collaboratively compute a function of their private inputs. 
\begin{itemize}
    \item $\textsf{MPHE.KeyGen}(\text{Params})\rightarrow (pk_{i},sk_{i})$ :Each party generates a pair of plain HE public key $pk_{i}$ and secret key $sk_{i}$.
    \item $\textsf{MPHE.DKeyGen}(\{pk_{i}\}_{i=0}^{N-1})\rightarrow cpk$ :It returns a collective public key $cpk$, for a set of public keys $\{pk_{i}\}$. Its corresponding secret key $csk$ is the aggregation of $\{sk_{i}\}_{i=0}^{N-1} $.
    \item $\textsf{MPHE.Enc}(cpk, m)\rightarrow ct$ :It returns a ciphertext $ct$ encrypting message $m$ using the collective public key $cpk$.
    \item $\textsf{MPHE.Eval}(cpk,\{ct_{i}\},f)\rightarrow ct_{f}$ :It takes as input the common public key $cpk$, a set of ciphertexts $\{ct_{i}\}$ all encrypted under $cpk$ and a function $f$, returns an evaluated ciphertext $ct_{f}$. Function $f$ includes addition \textsf{Add}, multiplication \textsf{Mul} and rotation \textsf{Rot}.
    \item $\textsf{MPHE.PubKeySwich}(ct,sk_i,tpk)\rightarrow h_i$: It takes as input a ciphertext ct encrypted under $cpk$, a secret key $sk_i$, where $csk=\sum_{i=0}^{N-1}sk_i$ associated with the corresponding public key $cpk$, and a public key $tpk$, outputs an intermediate result $h_i$, such that the aggregate ciphertext $ct' = \sum_{i=0}^{N-1}h_i$ can be decrypted using the corresponding secret key of $tpk$. 
    \item $\textsf{MPHE.Agg-Dec}(\{h_i\}_{i=0}^{N-1},tsk)\rightarrow m$: It takes as input the secret key $tsk$ and the set of intermediate results $\{h_i\}_{i=0}^{N-1}$. Initially, it aggregates $\{h_i\}_{i=0}^{N-1}$ to form a ciphertext $ct'$, and then decrypts $ct'$ to get message $m$ with target secret key $tsk$.
    \item $\textsf{MPHE.Reconstruct}(ct, \textsf{sk}_{i})\rightarrow pd_{i}$ : It takes as input a publicly known ciphertext $ct$, party's secret key $\textsf{sk}_{i}$, outputs a partial decryption $pd_{i}$.
    \item $\textsf{MPHE.Dec}(\{pd_{i}\}_{i=0}^{N-1},ct)\rightarrow m'$: It takes as input a set of partial decryption $\{pd_{i}\}_{i=0}^{N-1}$ and the corresponding publicly known ciphertext $ct$. By aggregating the partial decryption, it outputs message $m'$.
\end{itemize}
The scheme has the property as follows:

\noindent \textsf{Homomorphism}. The decryption result of ciphertext $ct = \textsf{MPHE.Eval}(cpk,\{ct_1,ct_2)\},f)$ should be $f(m_1,m_2)$, where $ct_1 = \textsf{MPHE.Enc}(cpk, m_1)$, $ct_2 = \textsf{MPHE.Enc}(cpk, m_2)$.

\subsection{Secure Aggregation Using MPHE}\label{sec:secure-aggregation}
 Secure aggregation is a concept primarily used in horizontal federated learning. It typically involves encrypting the updates $g_i$ from each client device in the system. These encrypted updates are then sent to the central server, which performs aggregation without being able to see the individual contributions. The aggregated result is then decrypted as $g$ where $g$ is the summation of each individual contribution such that $g=\sum g_i$, for the purpose of updating the global model. Secure Aggregation is a practical application of MPHE. In Hosseini et al. \cite{9682053}, it introduces a secure aggregation protocol for horizontal federated learning based on MPHE. Specifically, this protocol involves three phases as the following. 

 \paragraph{Setup Phase} Clients and the central server collaborate to generate common public key $cpk$. This is facilitated by the clients executing \textsf{MPHE.KeyGen} function, while the server performing \textsf{MPHE.DKeyGen}. 
 \paragraph{Aggregation Phase} Each client performs \textsf{MPHE.Enc} taking as input the local update $g_i$ and $cpk$, and outputs ciphertext $ct_i$.  Ciphertexts are transmitted to the server. 
 Upon receiving the set of ciphertexts $\{ct_i\}$ from the clients, the server computes $ct = \sum ct_i$, which represents the ciphertext of the aggregated updates. 
 \paragraph{Decryption Phase}Following this, the server and clients collaborate to decrypt $ct$. Each client contributes to this process by providing a partial decryption $pd_i$ generated using the \textsf{MPHE.Reconstruct} fucntion. Subsequently, the server aggregates these partial decryptions using the $\textsf{MPHE.Dec}$ function.


\subsection{Prior Work: POSEIDON} \label{sec:poseidon}
\begin{figure}[h]
    \centering
    \includegraphics[scale=0.42]{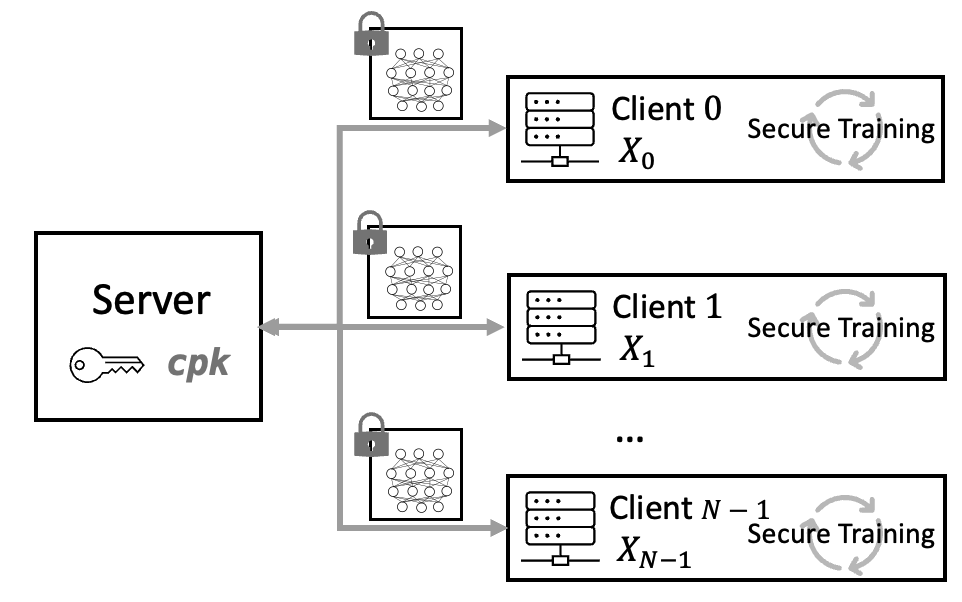} 
    \caption{POSEIDON's structure}
    \label{fig:poseidon}
\end{figure}

POSEIDON \cite{sav2020poseidon} is a secure neural network training protocol for horizontal federate learning. As shown in Figure~\ref{fig:poseidon}, in POSEIDON, the encrypted global weights will be transmitted to clients from the server. Each client trains the model based on its local data with SGD and collaboratively updates the global model. POSEIDON's main objective is to protect the global model parameters and the local data of each party. They make use of Multiparty Homomorphic Encryption CKKS scheme \cite{MPHE} to achieve such an objective. When the client receives the weights encrypted under $cpk$, it will perform homomorphic evaluation of neural network training between the ciphertext of the weights and the plaintext of the packed local data. The weights are kept encrypted all the time and all the intermediate training result are kept in the format of ciphertext, protecting the confidentiality of the model and training data. At the end of each communication round, the encrypted gradient will be transmitted to the server for aggregation and global model update.

POSEIDON is also applicable for secure inference once the model is  trained and the user possesses a complete set of features. In this scenario, the user downloads the encrypted model and conducts an encrypted inference process, akin to the feedforward procedure performed by each client during training. This approach enables the user to obtain the encrypted prediction result. Decryption of this result necessitates the involvement of all clients who contributed to  $cpk$. These clients provide their respective partial decryptions by executing \textsf{MPHE.Reconstruct}. Then the user aggregates the partial decryptions by executing \textsf{MPHE.Dec} to obtain the prediction result. While this framework is appealing, it requires transmission of the encrypted weights from the server to client nodes, which may not be desirable in practice. This also increases the computational load on client nodes during inference. It is also crucial to acknowledge that this method cannot be straightforwardly adapted to scenarios where data is vertically partitioned among clients.

\section{System Overview}
\subsection{System Setup}
We aim to develop privacy-preserving neural network inference protocol for vertical partitioned data. We focus on the setting as Figure~\ref{fig:setting} shows. In this setting, there are $N$ client parties, one server party, and one coordinator. Each of the client owns subset of features $\mathbf{X}_i$. The complete feature set $\mathbf{X}$ is the concatenation of subsets of features $\mathbf{X} = [\mathbf{X}_0||\mathbf{X}_1||...||\mathbf{X}_{N-1}]$. We assume that there are shared record identifiers among clients' local datasets, such as names, dates of birth, or universal identification numbers. The server party is deployed with a well pretrained neural network $\mathbf{F}$ in the encrypted form. At the start of the inference, a query is transmitted to the coordinator, specifying which sample input $\mathbf{X}$ should be used for the inference. In the end of inference, the coordinator will obtain the prediction result $\mathbf{y} = \mathbf{F}(\mathbf{X})$. 

\subsection{Privacy Goals and Threat Model}
Our main goal is to enable all the parties to collaboratively compute the prediction result of the deployed model with subset of features provided by client parties as input in a privacy-preserving way:
\begin{enumerate}
    \item The client $i$ should not leak $\mathbf{X}_i$ to any other party.
    \item The server party should not expose the weights of the deployed neural network $\mathbf{F}$.
    \item Any party should not learn any intermediate prediction result.
    \item Any party except for the coordinator should not learn the final prediction result.
\end{enumerate}

Each client can get access to the global information including: 1) Complete feature dimension such as height and width 2) The structure information of neural network. We also allow clients to know the indexes for their own local features.

 We assume all the parties to be honest-but-curious, that is, the party will not deviate from the protocol but will try to learn information about private inputs from other honest party. We assume dishonest majority of passive parties, that is up to $N-1$ parties can collude to learn other honest parties’ features by sharing their inputs and observations with dishonest parties. 
\section{The Protocol}
The protocol consists of two phases as discussed below. 
\subsection{Setup phase}
In the setup phase, all the parties in the system collaborate to generate two sets of keys: a common public key $cpk$ for feature encryption and a target key pair $(tpk, tsk)$ for secure decryption. The setup phase is only performed once at the beginning of the protocol. 
The generation of the two keys proceeds as follows:
\begin{enumerate}
    \item Each client $i$ executes \textsf{MPHE.KeyGen} to obtain a plain HE key pair $(pk_i,sk_i)$ and transmit $pk_i$ to the server.
    \item The server aggregates the public key from clients to generate a common public key by $cpk=\sum_{i=0}^{N-1}pk_i$.
    \item The server broadcasts the common public key $cpk$ to all clients.
    \item The coordinator generates HE key pair $(tpk,tsk)$ and sends target public key $tpk$ to all the clients.
\end{enumerate}
\begin{proposition}\label{prop1}
    One of the output of the setup phase, $cpk$, is a public key corresponding to secret key  $csk = \sum_i sk_i  \label{eq:sum}$.
\end{proposition}
  This step provides each client with a common public key $cpk$, a secret share $sk_i$ of the common secret key, and the target public key $tpk$, corresponding to $tsk$ held by only the coordinator. It is important to note that the secret key $csk$, corresponding to $cpk$, is not known to any party and the collective knowledge, that is needed to utilize it for decryption, exists within $N$ clients.


\subsection{Inference Phase}
\subsubsection{Secure Aggregation for Input Concatenation}

\textbf{Packing for Vertical Partitioned Data.} \label{pack-vfl}
When the input to a machine learning model is vertically partitioned, a straightforward method to form a complete input is to concatenate the portions. However, when each portion is encrypted, the concatenation becomes complex. We have developed a packing scheme, referred to as 
\textsf{V-Pack}, specifically tailored for input data that is vertically partitioned. This scheme enables the formation of encryption for the complete input by using the encryptions of processed portions as input through \textsf{V-Pack}. The key insight for packing of vertical partitioned data is to pad the subset features with $\mathbf{0}$ to be the complete sample and utilize the alternating packing scheme introduced in Protocol 3 (line 3 to line 6) in POSEIDON \cite{sav2020poseidon} to get the ciphertext of subset features. In this way, the server can collect the ciphertexts from clients, then form the ciphertext of complete features by simply adding the ciphertexts from clients together without knowing clients' identities. 
\begin{figure}[h]
    \centering
    \includegraphics[scale=0.63]{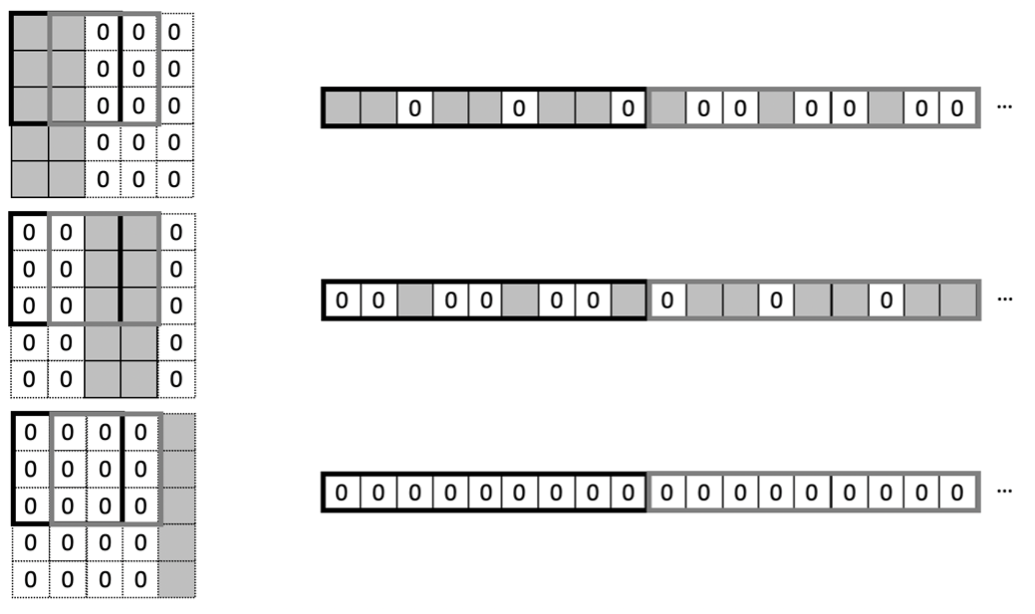} 
    \caption{Packing scheme for vertical partitioned input.}
    \label{fig:pack-conv-vfl}
\end{figure}

In details, each client packs its respective subset features.
We assume a single sample is partitioned by column index set $c=\{c_0,c_1,...c_{N-1}\}, 0 \leq c_i \leq F_W$.  Each party knows the indexes of the columns it can get access to. For $c=\{c_0,c_1,...c_{N-1}\}, 0 \leq c_i < F_W$. Each client knows the indexes of the columns they can access to. For example, client $i$ gets access to $\mathbf{X}_i = \mathbf{X}[c_i : c_{i+1}]$ while client $N-1$ accessing to $\mathbf{X}_0= \mathbf{X}[c_{N-1}:F_W]$. We assume there is no overlap between the column indexes each client can access to. Each party first completes the sample with $\mathbf{0}$, and follows the alternating packing scheme to preprocess the padded sample and then encrypt it with common public key $cpk$. 
Figure~\ref{fig:pack-conv-vfl} demonstrates a sample partitioned column-wise between three clients - with the first client holding the initial two columns, the second managing the third and fourth columns, and the third with the last column. Additionally, Figure~\ref{fig:pack-conv-vfl} also shows how each client pads their subset and subsequently preprocesses the padded sample.

\textbf{Secure Aggregation for Encrypted Concatenation.} The encrypted concatenation can be easily achieved via secure aggregation described in Section~\ref{sec:secure-aggregation}. In details, each client preprocesses its respective $\textbf{X}_i$ through \textsf{V-Pack} to get vector $\textbf{x}_i$, which is then encrypted as $ct_i$ using $cpk$. The ciphertexts are transmitted to the server. After receiving ciphertexts $\{ct_i\}$ from clients, the server compute $ct = \sum_{i=0}^{N-1}ct_i$ to form a ciphertext $ct$ of the vectorized complete features $\mathbf{X}$ as the input of the encrypted model.

\subsubsection{Inference and Decryption}

The MPHE scheme allows the server to perform neural network evaluation between the ciphertext of features and the ciphertext (or plaintext) of weights while keeping all the intermediate results encrypted. This procedure employs cryptographic neural network operations introduced in Section V.B and Section V.C in POSEIDON \cite{sav2020poseidon}. At the end of neural network evaluation,  the server holds the ciphertext of the prediction result $ct_y$. 

The server delivers $ct_y$ to all clients in the system for decryption. Each client first performs $\textsf{MPHE.PubKeySwitch}$ to obtain an intermediate decryption result $h_i$ using a secret share $sk_i$ of the common secret key and the target public key $tpk$. $h_i$ is sent to the coordinator for final decryption. After receiving $\{h_i\}_{i=0}^{N-1}$, the coordinator performs \textsf{MPHE.Agg-Dec} taking as input the set of $\{h_i\}$ and the target secret key $tsk$ to obtain the prediction result $y$.


\subsection{Privacy Analysis}
Recall the privacy goal of our protocol. We want the client $i$, the server and the coordinator to learn nothing about data $\mathbf{X}_j$ of client $j$ for $i \ne j$. We also want that any client and the coordinator should learn nothing about the deployed model weights $\mathbf{F}$. Additionally, we aim that the clients and the server should learn nothing about the final prediction result, and none of the parties should learn any intermediate results.
In this section, we prove that we have achieved the above-stated goal.
\begin{theorem}\label{mphe-theo}
 For any subsets of at most colluded $N-1$ clients by the adversary, for any two messages $m_1$ and $m_2$, no adversary has an advantage (better than $1/2$ chance) in distinguishing between distributions $\textsf{MPHE.Enc}(\textsf{cpk}, m_1)$ and \textsf{MPHE.Enc}(\textsf{cpk},$m_2$).    
\end{theorem}

After the setup phase, each client holds a common public key $cpk$, a secret share $sk_i$, and a target public key $tpk$. For each inference, each client packs the respective subset features according to the structure information of the deployed model but without knowing the exact parameters, and encrypts the packed data with $cpk$. Note that unlike VFL frameworks such as \cite{xu2021fedv, hardy2017private}, the encryption by each client only takes as input the respective data subset and public key $cpk$. Hence, each client cannot get access to the model deployed weights. The ciphertext of data subset is sent to the server for concatenation and evaluation. During the process, any intermediate results will be the encrypted form and will not leak any information or be decrypted unless $N$ clients collude together based on Proposition~\ref{prop1} and Theorem~\ref{mphe-theo}. Thus, the client $i$ and the server won't be able to learn other client's data $\mathbf{X}_j$ or any intermediate processing results.

\textbf{Importance of using target key pair for decryption.} 
The distribute decryption proceeds as follows: The server broadcasts the ciphertext of prediction result to all the clients. Utilizing their respective $sk_i$ and the shared $tpk$, each client executes the \textsf{MPHE.PubKeySwitch} function, generating an intermediate result $h_i$. The intermediate results from clients will be first aggregated and then decrypted by the coordinator with $tsk$. It is important to note that \textsf{MPHE.PubKeySwitch} and \textsf{MPHE.Agg-Dec} are useful when the clients know only a public key ($tpk$) for the secret key ($tsk$) of the decryptor, who is the coordinator. Since it necessitates only public input from the clients, the \textsf{MPHE.PubKeySwitch} and \textsf{MPHE.Agg-Dec} functions facilitate the coordinator, one not included in the input access-structure, to acquire a decryption of the output without requiring private communication channels with the clients. In other words, the coordinator party can be user's personal devices such as phone or laptop. This decryption process ensure that only the user will know the actual value of prediction result, achieving end-to-end security.

A simpler decryption method that does not utilize the target key pair, as detailed in Section~\ref{sec:secure-aggregation} , involves clients using 
\textsf{MPHE.Reconstruct} with their $sk_i$ to generate partial decryptions. The coordinator then aggregates these using $\textsf{MPHE.Dec}$ to reveal the prediction result.
However, this approach presents a security risk: an eavesdropper could potentially intercept the partial decryptions communicated between the coordinator and the clients. Since the aggregation of these partial decryptions doesn't require a secondary decryption, the prediction result is vulnerable to leakage. Therefore, the use of the target key pair, along with the associated \textsf{MPHE.PubKeySwitch} and \textsf{MPHE.Agg-Dec} function, is necessary to ensure the confidentiality of the private information.

\section{Experiments}
In this section, we empirically evaluate the protocol in terms of the execution time, the communication cost, the scalability performance.
\subsection{Setup}
We implemented the proposed protocol using Lattigo \cite{lattigo} library. we employ the CKKS scheme \cite{cheon2017homomorphic} as the foundational encryption scheme. The size of the ciphertext space, $q$, is configured to be a 60-bit prime number. We set the order of the polynomial, $n$, to 8192 and adjust the remaining parameters to ensure 256-bit security, in accordance with \cite{albrecht2021homomorphic}. The net package of Golang is used to build the communication system. The experiments were carried out on 4 instances, each with 8 cores, 16GB RAM. We evaluate the protocol on a CNN model with 2 convolution layers and a total 1198 parameters . The input data used in the experiment is MNIST figure.
\subsection{Experimental Results}
\begin{figure}[h]
     \centering
     \begin{subfigure}[b]{0.25\textwidth}
         \centering
         \includegraphics[scale=0.24]{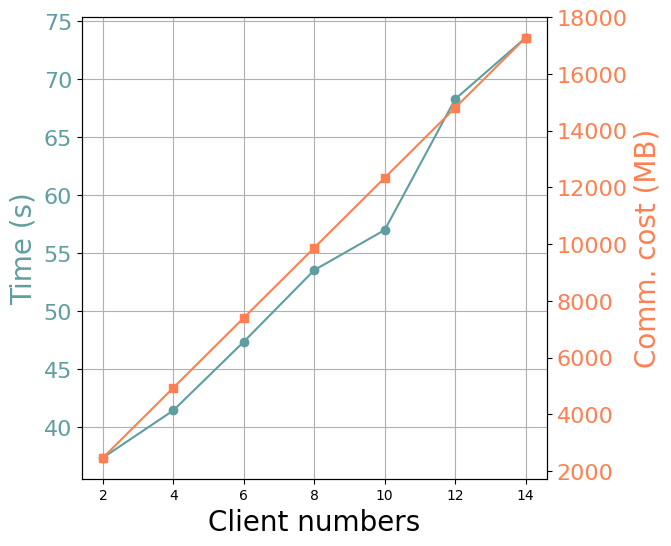}
        \caption{Setup}
        \label{}
     \end{subfigure}\begin{subfigure}[b]{0.25\textwidth}
        \centering
        \includegraphics[scale=0.24]{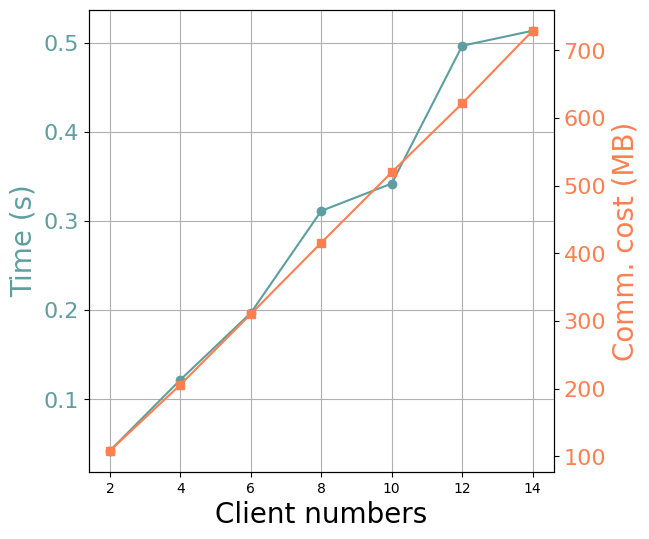}
        \caption{Encrypted Concatenation}
        \label{}
    \end{subfigure}
    \begin{subfigure}[b]{0.25\textwidth}
        \centering
        \includegraphics[scale=0.24]{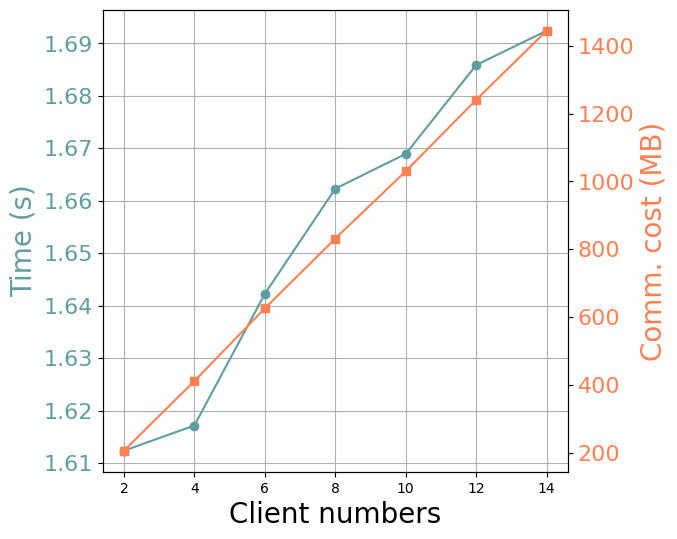}
        \caption{Distribute Decryption}
        \label{}
    \end{subfigure}
    \caption{Time and communication cost of some steps vs. The number of clients }
    \label{fig:numofclients}
\end{figure}

\paragraph{Execution Time and Communication Cost}
In this section, we assess the protocol's execution time and communication costs across each phase, as detailed in Table~\ref{tab:exe-comm}. The input data is divided among three clients. We specifically examine two scenarios: one where the server's model is in plaintext or in ciphertext. It's important to note that the communication costs in this step are from the transmission of ciphertexts from the clients, which is independent of the server's model format. Consequently, the communication costs for these two scenarios remain identical. Regarding time efficiency, inference with ciphertext model requires slightly more time than with plaintext model. This additional time can be justified by the enhanced security it offers to the model owner, since the server knows nothing about the model parameters except for its structure if the deployed model is in ciphertext. Additionally, we observe that the main cost is from the setup phase but is a one-time cost, since the setup phase is a one-time execution for the same set of clients, coordinator, and server and all the keys can be reused. The time of inference depends on the type (ciphertext or plaintext) and structure of the deployed model and is independent of number of clients, while other costs increase as the number of clients increases. 
\begin{table}[]
\centering
\caption{Execution time and Communication cost of each phase}\label{tab:exe-comm}
{\scriptsize
\begin{tabular}{ccc}
\toprule
                                & Time cost (s) & Communication cost (MB) \\\midrule
Key Generation                  & 54.45         & 2988.24                 \\
Inference with plaintext model  & 30.64         & 154.2                   \\
Inference with ciphertext model & 31.97         & 154.2                   \\
Distribute Decryption           & 7.34          & 308.4                  \\\bottomrule
\end{tabular}}
\end{table}
\paragraph{Scalability}
In this section, we focus on the execution time and communication cost of key setup, the encrypted concatenation and the distribute decryption to assess scalability performance, as the cost of the three steps is affected by the number of clients. The number of clients ranges from 2 to 14 as we consider the data is vertically partitioned, which typically involves a limited number of data holders. Figure~\ref{fig:numofclients} shows that the time and the communication cost of these steps linearly increase as the number of clients increases. 

\section{Conclusions}
We propose an end-to-end N-party secure inference protocol for vertically partitioned data. The protocol ensures stronger security of data and model via Multiparty HE scheme. By integrating a novel packing scheme, the protocol enables encrypted data concatenation. It upholds the confidentiality of prediction results and facilitates private collaborative analysis or prediction without compromising data or model privacy. 

\balance
\bibliographystyle{IEEEtran}
\bibliography{main}
\clearpage

\end{document}